\input phyzzx.tex

\twelvepoint


\def\A{{\cal A}}
\def\bA{{\bar {\cal A}}}
\def\S{{\cal S}}
\def\bS{{\bar {\cal S}}}
\def\U{{\cal U}}
\def\bU{{\bar {\cal U}}}
\def\bL{{\bf \Lambda}}

\def\bz{{\bar z}}
\def\bs{{\bar s}}
\def\ba{{\bar a}}
\def\bu{{\bar u}}


\REF\Witten{E. Witten, Nucl. Phys. {\bf B500} (1997) 3, hep-th/9703166}
\REF\SW{N. Seiberg and E. Witten, Nucl. Phys. {\bf B426} (1994) 19, 
hep-th/9407087}
\REF\KLVW{A. Klemm, W. Lerche, C. Vafa and N., Warner,
Nucl. Phys. {\bf B477} (1996) 746, hep-th/9604034}
\REF\three{P.S. Howe, N.D. Lambert and P.C. West, {\it The Threebrane Soliton
of the M-fivebrane}, hep-th/9710033}
\REF\HLW{P.S. Howe, N.D. Lambert and P.C. West, {\it Classical M-Fivebrane
Dynamics and Quantum $N=2$ Yang-Mills}, hep-th/9710034}
\REF\LW{N.D. Lambert and P.C. West, {\it Gauge Fields and M-Fivebrane
Dynamics}, hep-th/9712040}
\REF\pert{P. Howe, K. Stelle and P. West, Phys. Lett. {\bf 124B} (1983) 55;
see also
P. West, {\it in  Proceedings of the 1983
Shelter Island II Conference on Quantum Field Theory and Fundamental
Problems of Physics}, edited by R. Jackiw, N. Kuri , S. Weinberg and
E. Witten (M.I.T. Press); P.S. Howe, K.S. Stelle
and P.K. Townsend,  Nucl. Phys. {\bf B236} (1984) 125,
M. Grisaru and W. Siegel, Nucl. Phys {\bf B201} (1982)292.}
\REF\inst{N. Dorey, V. V. Khoze and M. P. Mattis, Phys. lett. {\bf B388}
(1996) 324, hep-th/9607066; Phys. Rev.  {\bf D54} (1996) 2921; Phys. Rev. 
{\bf D54} (1996) 7832, hep-th/9607202; 
K. Ito and N. Sasakura, Phys. Lett. {\bf B382} (1996) 95, hep-th/9602073; 
Nucl. Phys. {\bf B484} (1997) 141, hep-th/9608054; 
A. Yung, Nucl. Phys. {\bf B485} (1997) 38, hep-th/9605096;
H. Aoyama, T. Harano, M. Sato and S. Wada, Phys. lett. {\bf B388} (1996) 331, 
hep-th/9607076;
T. Harano and M. Sato, Nucl.Phys. {\bf B484} (1997) 167, hep-th/9608060}
\REF\GSW{R. Grimm, M. Sohnius and J. Wess, Nucl. Phys. {\bf B133} (1978) 275}
\REF\M{L. Mezincescu, JINR report P2-12572 (1979)}
\REF\vP{This follows a discussion between A. van Proeyen and P. West.}
\REF\STG{G. Sierra and P.K. Townsend, {\it in Proceedings, Supersymmetry and
Supergravity 1983}, ed. B. Milewski, World Scientific, 1983; 
S.J. Gates Jr., Nucl. Phys. {\bf B238} (1984) 349}

\pubnum={KCL-TH-98-01  \cr hep-th/9801104}
\date{January 1998}

\titlepage

\title{\bf $N=2$ Superfields and the M-Fivebrane}

\centerline{N.D. Lambert}

\centerline{and}

\centerline{P.C. West\foot{lambert, pwest@mth.kcl.ac.uk}}
\address{Department of Mathematics\break
         King's College, London\break
         England\break
         WC2R 2LS\break
         }

\abstract

In this paper we provide a manifestly $N=2$ supersymmetric formulation
of the M-fivebrane in the presence of threebrane solitons. The 
superspace form of four-dimensional effective equations for the threebranes 
are readily  obtained and yield the complete
Seiberg-Witten equations of motion for $N=2$ super-Yang-Mills. 
A particularly simple derivation is given by introducing
an $N=2$ superfield generalisation of the Seiberg-Witten differential.

\endpage


\chapter{Introduction}

One of the most interesting recent developments is the connection between
classical brane dynamics and quantum field theories. 
This work was initiated by Witten [\Witten] who studied a configuration 
of NS-fivebranes and D-fourbranes in type IIA string theory, 
for which the low energy effective
action is a four-dimensional $N=2$ Yang-Mills theory.  
From the M theory point of view this configuration is a single
M-fivebrane with a complicated set of self-intersections and it was argued 
in [\Witten] that it is in fact wrapped on a 
Riemann surface $\Sigma$. This identification 
provides a systematic explanation of the origin of 
the auxiliary curve in the Seiberg-Witten solution of $N=2$ Yang-Mills [\SW] 
for various gauge groups and matter content and also illuminates many 
qualitative features of quantum Yang-Mills theories. A similar role for the
Riemann surface also appeared in [\KLVW] when considering Calabi-Yau
compactifications of type II strings.

In [\three]
the worldvolume soliton solution for the intersection of two M-fivebranes
(or self-intersection of a single M-fivebrane) was found. 
The Riemann surface then appears
naturally as a consequence of the Bogomoln'yi condition. It was further
shown in [\HLW,\LW] that the classical effective action for this soliton  
could be calculated from the M-fivebrane's equations of motion and 
the lowest order terms are precisely the full Seiberg-Witten effective 
action for quantum $N=2$ Yang-Mills theory.  
Thus the classical M-fivebrane dynamics contains the entire quantum effects 
of low energy $N=2$ $SU(2)$ Yang-Mills theory. 
This correspondence not only predicts the correct known
perturbative contributions to the $\beta$-function [\pert] but also an 
infinite number of instanton corrections, of which only
the first two have been found by explicit calculation [\inst]. 

In this paper we will restrict our attention to
the case of two threebranes whose centre of mass is fixed. 
However our work can be readily extended to the more general case. 
In the corresponding type IIA picture one sees two NS-fivebranes with two 
D-fourbranes  suspended between them so that the low energy 
effective action is 
$N=2$ $SU(2)$ Yang-Mills [\Witten]. 
We provide a complete $N=2$ superspace 
proof of the equivalence between the low energy
motion of threebranes in the M-fivebrane and the low energy Seiberg-Witten
effective action for quantum $N=2$ $SU(2)$ 
super-Yang-Mills, extending  the results 
of [\HLW,\LW] to include the  fermionic zero modes. However our primary
motivation for this work is to see what insights and simplifications
can be gained into 
the M-fivebrane/Seiberg-Witten correspondence from $N=2$ superspace, which
provides a geometrical unification of  all of the zero modes.
Indeed many of subtleties which appear in the purely bosonic formulations
can be clearly observed and understood in  $N=2$ superspace. 


\chapter{$N=2$ and $N=1$ Superfields}

Let us first recall how  the $N=2$ chiral effective action  decomposes  
in terms of $N=1$ superfields. In this paper $N=2$ superfields are
denoted
by calligraphic or bold faced letters, $N=1$ superfields by 
upper case letters and $N=0$ fields by lower case letters. 
$N=2$ Yang-Mills theory is described by a chiral superfield $\A$ [\GSW]
$$
{\bar D}_{\dot B}^i \A=0\ ,\
\eqn\chrialdef
$$
which also satisfies the constraint
$$
D^{ij} \A = {\bar D}^{ij}{\bA}\ ,
\eqn\constraint
$$
where $D^{ij} = D^{Ai}D_{A}^{\ j}$ and ${\bar D}^{ij} = 
{\bar D}^{{\dot A}i}{\bar D}_{\dot A}^{\ j}$. 
This constraint then ensures that the vector
component obeys the Bianchi identity and that the auxiliary field is real. 
The chiral effective action may then be written as
$$
S = {\rm Im}\int d^4x d^4\theta F(\A) \ .
\eqn\YMaction
$$
The solution to the above constraints in the Abelian case is of the 
form [\M]
$$
\A = {\bar D}^4D^{ij}{\cal V}_{ij}\ ,
\eqn\Vdef
$$
where ${\cal V}_{ij}$ is an unconstrained superfield of mass dimension $-2$. 
Varying the action \YMaction\ with respect to ${\cal V}_{ij}$ yields 
$$
{\rm Im}\int d^4 x d^4\theta {\bar D}^4 D^{ij}\delta {\cal V}_{ij}
{dF(\A)\over d\A}
= 16{\rm Im} \int d^4 x d^8\theta  
\delta {\cal V}_{ij}D^{ij}\left({dF(\A)\over d\A}\right)\ ,
\eqn\variation
$$
and hence we find the equation of motion to be 
$$
D^{ij}{dF\over d\A} = {\bar D}^{ij} {d {\bar F}\over d \bA}\ .
\eqn\YMeqofm
$$
We observe that if we write $\A_D = dF/d\A$ and $\A$ as a doublet then the
constraint \constraint\ and the equations of motion \YMeqofm\ can be
written as
$$
D^{ij}\left(\matrix{\A_D\cr \A \cr}\right) 
= {\bar D}^{ij}\left(\matrix{{\bA}_D\cr {\bA} \cr}\right)\ .
\eqn\system
$$
Clearly this system has a set of equations invariant under
$$
\left(\matrix{\A_D\cr \A \cr}\right)\rightarrow 
\Omega \left(\matrix{\A_D\cr \A \cr}\right)\ ,
\eqn\sym
$$
where $\Omega \in SL(2,{\bf Z})$. Thus an $SL(2,{\bf Z})$ symmetry is
naturally realised in $N=2$ superspace [\vP].

For a free theory $F= iA^2$ and we find the equation of motion and constraint
imply $D^{ij} \A = {\bar D}^{ij} \A= 0$. At $\theta^{Ai}=0$ this equation 
sets the auxiliary field to zero and ensures the correct equation of motion
for the Yang-Mills theory.

We now solve the $N=2$ superspace constraints in terms of $N=1$ superfields.
Let us label the two superspace Grassmann odd coordinates which occur in the
$N=2$ superspace as $\theta^{A1}=\theta^A$ and $\theta^{A2}=\eta^{A}$ and
similarly for their conjugates. We also introduce the derivatives
$D_A =D_A^{\ 1}$ and $\nabla_A=D_{A}^{\ 2}$ for which $D^2=D^AD_A$, 
${\bar D}^2={\bar D}^{\dot A}{\bar D}_{\dot A}$
and similarly for $\nabla^2$ and ${\bar \nabla}^2$.  
We associate the coordinates $\theta^A$
and ${\bar \theta}^{\dot A}$ with those of the $N=1$ superspace, which we will
keep manifest. The $N=2$ superfield $\A$ can be decomposed into the
$N=1$ superfields $A=\A\mid_{\eta=0}$, 
$W_A = \nabla_A\A\mid_{\eta=0}$ and $G = -{1\over2}\nabla^2\A\mid_{\eta=0}$.
These $N=1$ superfields are also $N=1$ chiral since $\A$ satisfies \chrialdef;
$$
{\bar D}_{\dot B}A = 0\ ,\quad  {\bar D}_{\dot B}W_C = 0\ .
\eqn\onechiral
$$
It remains to solve the constraint \constraint .
Taking $i=1,\ j=2$ we find
the constraint  $D^BW_B= {\bar D}^{\dot B} {\bar W}_{\dot B}$.
Taking $i=j=2$ we find that
$$
\nabla^2 \A = {\bar D}^2{\bA}\ ,
\eqn\twochiral
$$
which implies $G=-{1\over 2}{\bar D}^2 {\bar A}$ and so $G$ 
is not an independent superfield.
 
We can now evaluate the action \YMaction\ in terms of $N=1$ superfields
[\STG]
$$\eqalign{
S  & =
{\rm Im}\int d^4 x d^2 \theta d^2\eta F \ ,\cr
&= {\rm Im}\int d^4 x d^2 \theta\left\{-{\nabla^2{\A}\over 4}
{dF\over d\A} 
 - {1\over4} {\nabla}^{ B} \A {\nabla}_{B} \A    
{d^2F \over d\A^2}\right\}_{\eta=0}\ ,\cr
&= {\rm Im}\int d^4 x d^2 \theta\left\{-{\bar D^2{\bar A}\over 4}
{dF\over dA} 
- {1\over4}{d^2F\over dA^2} {W}^{B}{W}_{B}\right\}\ ,\cr
&= {\rm Im}\left\{\int d^4 x d^4\theta{\bar A}{dF\over dA}
- {1\over4}\int d^4 x d^2\theta {d^2F\over dA^2}{W}^{B}{W}_{B}\right\} \ .\cr}
\eqn\oneaction
$$
This is easily recognised as the standard form 
for the action of an $N=2$
Abelian Yang-Mills multiplet written in $N=1$ superfields.

At this point we remark on an important notational oddity. Consider the
coefficient of 
$D_AW_B|_{\theta=0}$. Following standard conventions 
the generators of the four-dimensional Lorentz group $\sigma^{\mu\nu}$ 
satisfy
$\sigma^{\mu\nu} = -{i\over 2}\epsilon^{\mu\nu\rho\lambda}
\sigma_{\rho\lambda}$. Therefore 
$D_AW_B |_{\theta=0}= \sigma_{AB}^{\mu\nu}(F_{\mu\nu} -i\star F_{\mu\nu})$, 
where 
$F_{\mu\nu}$ is the curl of a four-dimensional gauge field. 
Now the M-fivebrane field content contains a self-dual three form 
$H$ and  one can check that 
$F_{\mu\nu} -i\star F_{\mu\nu}$ must come  
from the $H_{\mu\nu\bz}$ component of $H$ (this was also shown in [\LW]). 
Thus one sees that the lowest component of $A$ must be $\ba(\bz)$. 
In this paper then unbarred superfields depend upon
$\bz$ and barred superfields depend on  $z$. 
This problem has its origins
in the two uses for the bar symbol; as complex conjugation on the Riemann
surface, or as Hermitian conjugation in $N=2$ superspace. We are choosing
the later definition to have precedence.


\chapter{The M-Fivebrane and Seiberg-Witten}

We refer the reader to [\HLW,\LW] for a detailed 
discussion of the threebrane and its
zero modes. The Riemann surface $\Sigma$ for two threebranes is given by  
$t^2 - 2(z^2 - u)t + \Lambda^4 = 0$, where $t = e^{-s}$, $z$ is a coordinate
of the surface and  $u$ is a modulus
related to the relative separation of the two threebranes.
From a supersymmetric point of view it is natural to take the  
complex scalar $\bs(\bz)$ to be the lowest component of a chiral 
$N=2$ superfield $\S$ with independent $N=1$ components
$$
\S\mid_{\eta=0} = S\ , \quad \nabla_A \S\mid_{\eta=0} = H_{A \bz}\ ,
\eqn\Scomponents
$$
with $S\mid_{\theta =0} = \bs$ and 
$D_AH_{B\bz}|_{\theta=0}=\sigma_{AB}^{\mu\nu}H_{\mu\nu\bz}$.
Similarly we promote the moduli $\bu$ of the Riemann surface to an
$N=2$ superfield $\U$ with the independent $N=1$  components
$$
\U\mid_{\eta=0} = U\ , \quad \nabla_A \U\mid_{\eta=0} = T_{A}\ ,
\eqn\Ucomponents
$$
and $U\mid_{\theta =0} = \bu$.
We can then define the $N=2$ superfields
$$
\A = \oint_{A}\S d\bz\ ,\quad \A_D = \oint_{B}\S d\bz\ ,
\eqn\AADdef
$$
where $A$ and $B$ are a basis of  one cycles of $\Sigma$. 
Given these definitions one can see that the $\eta^A$ components of $\A$  
and $\A_D$ are
$$\eqalign{
W_{A} &=\nabla_A \oint_{A}\S d\bz 
= \oint_{A} {\Lambda (U)} T_A 
= {dA\over dU} T_A\ ,\cr
W^D_{A} &=\nabla_A \oint_{B}\S d\bz 
= \oint_{B} {\Lambda (U)} T_A 
= {dA_D\over dU} T_A\ .\cr
}
\eqn\WWD
$$
Here $\Lambda  = {dS\over dU} d\bz$ is an $N=1$ superfield whose lowest
component is the anti-holomorphic one form  ${\bar \lambda}$ 
of the Riemann surface. Furthermore it follows from  
$$
H_{A\bz} = \nabla_A\S|_{\eta=0} 
= {dS\over dU}\nabla_A\U|_{\eta=0}
= {dS\over dU}T_A\ ,
\eqn\Teval
$$
and \WWD\ that
$$
H_{\mu\nu\bz} 
= \left({d{S}\over d{A}}\right)(F_{\mu\nu}-i\star  F_{\mu\nu})
= \left({dA\over dU}\right)^{-1}
(F_{\mu\nu}-i\star  F_{\mu\nu})\Lambda_{\bz}\ ,
\eqn\ansatztwo
$$
which agrees with the ansatz used in [\LW]
for the vector zero modes.

Now we wish to obtain a manifestly $N=2$ formulation of the equations of
motion for the threebranes of the M-fivebrane [\LW].
To this end we postulate the following equation
$$
{\cal E} \equiv D^{ij}\S - R^2\Lambda^4\partial_{\bz}
\left({D^{Ai}\S D_A^{\ \ j}\S\partial_z\bS\over 
1+R^2\Lambda^4|{\partial_{\bz}}\S|^2 } 
- {D^{{\dot A}i}\bS {\bar D}_{\dot A}^{\ \ j}\bS\partial_{\bz}\S\over 
1+R^2\Lambda^4|\partial_{\bz}\S|^2}\right)=0\ .
\eqn\susyeq
$$
First we take the $i=j=1$ component of \susyeq. This gives at $\eta=0$
$$
D^2 S - R^2\Lambda^4 \partial_{\bz}\left(
{D^{A}S D_A S\partial_z{\bar S}\over 
1+R^2\Lambda^4|{\partial_{\bz}}S|^2 } 
- {{\bar T}^{\dot A} {\bar T}_{\dot A}\partial_{\bz}S\over 
1+R^2\Lambda^4|\partial_{\bz}S|^2}
\right)=0 \ .
\eqn\ijone
$$
Next we act on \ijone\ with ${\bar D}^2$ and set the fermions to zero  to 
obtain
$$
{\bar D}^2D^2 S + 2R^2\Lambda^4 \partial_{\bz}\left(
{{\bar D}^{\dot B}D^{A}S {\bar D}_{\dot B}D_A S\partial_z{\bar S}\over 
1+R^2\Lambda^4|{\partial_{\bz}}S|^2} 
- {{\bar D}^{\dot B}{\bar T}^{\dot A} {\bar D}_{\dot B}{\bar T}_{\dot A}
\partial_{\bz}S\over  1+R^2\Lambda^4|\partial_{\bz}S|^2}
\right)=0 \ .
\eqn\ijtwo
$$
This equation can then be evaluated to give
$$
\partial_{\mu}\partial^{\mu}\bs 
- R^2\Lambda^4\partial_{\bz}\left(
{\partial_{\mu}\bs\partial^{\mu}\bs\partial_z s\over 
1+R^2\Lambda^4|{\partial_{z}}s|^2}
+ H_{\mu\nu z}H^{\mu\nu}_{\ \ z}
{\partial_{\bz}\bs\over 1+R^2\Lambda^4|{\partial_{z}}s|^2}\right)=0\ ,
\eqn\ijthree
$$
which is precisely the equation for the scalar zero modes obtained in [\LW],
provided that we rescale $H_{\mu\nu z}\rightarrow 4H_{\mu\nu z}$.

Now we take the $i=1$, $j=2$ component of \susyeq. At $\eta=0$ we find
$$
D^AT_A + R^2\Lambda^4\partial_{\bz}\left(
{D^A S T_A\partial_z{\bar S}\over 1+R^2\Lambda^4|\partial_{\bz}S|^2}
+ {{\bar T}^{\dot A}{\bar D}_{\dot A}{\bar S} \partial_{\bz}S
\over 1+R^2\Lambda^4|{\partial_{\bz}}S|^2} \right)=0 \ .
\eqn\ijfour 
$$
Next we act with $D_C{\bar D}_{\dot B}$ on \ijfour\ and set
the fermions to zero to obtain
$$
D_C{\bar D}_{\dot B}D^AT_A
-R^2\Lambda^4\partial_{\bz}\left(
{\partial_z {\bar S} {\bar D}_{\dot B}D^A S D_C T_A
\over 1+R^2\Lambda^4|\partial_{z}{\bar S}|^2}
- {\partial_{\bz}SD_C{\bar D}^{\dot A}{\bar S}
{\bar D}_{\dot B}{\bar T}_{\dot A}
\over 1+R^2\Lambda^4|\partial_{z}{\bar S}|^2}
\right)=0\ .
\eqn\ijfive
$$
Evaluating \ijfive\ we then arrive at the equation for the vector 
zero modes
$$
\partial^{\nu}H_{\mu\nu\bz} - R^2\Lambda^4\partial_{\bz}\left(
{\partial_z s\partial^{\nu}\bs H_{\mu\nu\bz}
\over 1+R^2\Lambda^4|{\partial_{z}}s|^2}
-{\partial_{\bz}\bs \partial^{\nu}s H_{\mu\nu z}
\over 1+R^2\Lambda^4|{\partial_{z}}s|^2}\right)=0\ ,
\eqn\ijsix
$$
which is again precisely the same equation that was found in [\LW].
Since the scalar and vector component equations agree with the correct
equations of motion, it follows that \susyeq\ describes the all zero modes
of a threebrane soliton in the M-fivebrane worldvolume.

Now that we have shown that \susyeq\ reproduces the M-fivebrane equations
of motion we can determine the equations of motion for the massless modes
of the threebranes solitons by reducing ${\cal E}=0$ 
over the Riemann surface. 
Thus we consider
$$
0 = \int_{\Sigma} {\cal E}d\bz\wedge {\bar \bL}\ ,
\eqn\reduce
$$
where, according to our conventions ${\bar \bL} = 
{d{\bS}\over d{\bU}}dz$ is an $N=2$ superfield whose lowest component is
the holomorphic one form $\lambda$.
Expanding out the terms in \susyeq\ one finds
$$
0 = D^{ij}\U I 
+ D^{Ai}\U D_{A}^{\ \ j}\U {dI\over d\U}
- D^{Ai}\U D_{A}^{\ \ j}\U J 
+ {\bar D}^{{\dot A}i}\bU {\bar D}_{\dot A}^{\ \ j}\bU K\ ,
\eqn\eqofmo
$$
where the $I,J$ and $K$ integrals
have appeared before in [\LW] where  their 
values have also been deduced; 
$$\eqalign{
I &\equiv \int_{\Sigma}{\bL}\wedge{\bar \bL} 
= {d\A\over d \U}{d\bA\over d\bU}({\tau} - {\bar \tau})\ ,\cr
J&\equiv R^2\Lambda^4\int_{\Sigma}\partial_{\bz}\left(
{\bL_{\bz}^2\partial_{z}\bS\over 1+R^2\Lambda^4\partial_{\bz} 
\S\partial_{z}\bS}\right)d\bz\wedge{\bar \bL} = 0\ ,\cr
K &\equiv R^2\Lambda^4 \int_{\Sigma}\partial_{\bz}\left(
{{\bar \bL}_{z}^2\partial_{\bz}\S\over 1+R^2\Lambda^4\partial_{\bz} \S
\partial_{z}\bS} \right)d\bz\wedge {\bar \bL}  
= -{d{\bar \tau}\over d{\bU}}\left({d{\bA}\over d{\bU}}\right)^2
\ ,\cr}
\eqn\IJK
$$
where $\tau(\U) = {d{\A}_D\over d\A}$.
The first integral is easily evaluated using the Riemann bilinear
identity, however in [\LW] the two non-holomorhpic integrals required a 
rather indirect
method to evaluate them. In the appendix to this paper we provide a 
direct proof of the above expressions for $J$ and $K$.

Multiplying by $({d\bA\over d\bU})^{-1}$ one can rewrite \eqofmo\ in the
simple form
$$
D^{ij}{\A}_D-{\bar D}^{ij}\bA_D-{\bar \tau}(D^{ij}\A -{\bar D}^{ij}\bA) = 0\ .
\eqn\eqofmotwo
$$
The real and imaginary parts of this equation are equivalent to
$$
D^{ij}\A = {\bar D}^{ij}\bA\ , \quad
D^{ij}{\A}_D = {\bar D}^{ij}{\bA}_D\ ,
\eqn\eqofmori
$$
respectively. Therefore if we introduce a function $F(\A)$ defined so that
$\A_D = {\partial F\over \partial\A}$ then we see
from \AADdef\ 
that these equations are precisely those of the Seiberg-Witten effective
theory for $N=2$ $SU(2)$ Yang-Mills with $N=2$ superspace action \YMaction. 

Finally let us consider the following $N=2$ 
superspace generalisation of the Seiberg-Witten differential,
$\bL_{SW}= \S d\bz$. The lowest  component of $\bL_{SW}$ 
is the Seiberg-Witten
differential ${\bar \lambda}_{SW}$ and its $\eta^A$ component is the form
$H_{A\bz}d\bz$. First we note that for either the $A$ or $B$ cycle
$$
\oint ({\cal E}d\bz - {\bar {\cal E}}dz)
= \oint (D^{ij}\bL_{SW}-{\bar D}^{ij}{\bar \bL}_{SW})\  ,
\eqn\cycle
$$
since the non-holomorphic terms in $\cal E$ collect into the form
$d(f-{\bar f})$. We can discard the integral of $d(f-{\bar f})$ over the $A$ 
or $B$ cycles as the function
$f(\S,\bS)$ is non-singular in a neighbourhood of these cycles, which are
therefore closed curves on $\Sigma$. 
A direct derivation of the Seiberg-Witten effective
equations of motion comes from evaluating
$$\eqalign{
0 & = \oint_A \left( {\cal E}d\bz - {\bar {\cal E}}dz\right)
 =  \oint_A\left(
D^{ij}\bL_{SW}-{\bar D}^{ij}{\bar \bL}_{SW}\right)\ ,\cr
0 & = \oint_B \left( {\cal E}d\bz - {\bar {\cal E}}dz\right)
 =  \oint_B\left(
D^{ij}\bL_{SW}-{\bar D}^{ij}{\bar \bL}_{SW}\right)\ ,\cr}
\eqn\quick
$$
which yields
$$
D^{ij}\left(\matrix{\oint_{A}{\bL_{SW}}\cr \oint_{B}\bL_{SW} \cr}
\right) = {\bar D}^{ij}\left(
\matrix{\oint_{A}{\bar \bL}_{SW}\cr \oint_{B}{\bar \bL}_{SW} \cr}
\right)\ .
\eqn\eqofm
$$
Clearly an $SL(2,{\bf Z})$ transformation on the $(A,B)$
cycles generates the  $SL(2,{\bf Z})$ transformation on $(A,A_D)$ discussed
in the previous section. We note that the condition 
$D^{ij}{\bf \Lambda}_{SW} =  {\bar D}^{ij}{\bar \bL}_{SW}$ is simply the
constraint \constraint\ applied to $\bL_{SW}$. 
Thus the Seiberg-Witten equations of motion can be obtained 
by imposing the $N=2$
superfield constraint \constraint\ on the generalised Seiberg-Witten
differential $\bL_{SW}$ and then integrating it over the 
cycles of the Riemann surface.

To make contact with the previous discussion we note that 
because $dz\wedge {\bar \bL}=0$ \reduce\ can be 
rewritten as  
$$\eqalign{
0 &= \int_{\Sigma} \left( {\cal E}d\bz - {\bar {\cal E}}dz\right)
\wedge {\bar \bL} \cr
&= \oint_B\left(
D^{ij}\bL_{SW}-{\bar D}^{ij}{\bar \bL}_{SW}\right)
\oint_A  {\bar \bL} 
- \oint_A\left(
D^{ij}\bL_{SW}-{\bar D}^{ij}{\bar \bL}_{SW}\right)
\oint_B {\bar \bL} \cr
&=(D^{ij}{\A}_D - {\bar D}^{ij}{\bA}_D){d\bA\over d\bU} 
- (D^{ij}\A - {\bar D}^{ij}\bA){d{\bA}_D\over d\bU}\ ,}
\eqn\redconst
$$
where we have applied the Riemann bilinear relation and again dropped the
total derivative terms. In this way we arrive immediately at equation 
\eqofmotwo, whose real and imaginary parts are \eqofm.

We would like to thank A. Pressley for discussions on Riemann surfaces.

Note Added

From the above superfield construction one sees a potential obstruction
to obtaining the Seiberg-Witten effective action from  a 
six-dimensional action. The chiral nature of $N=2$ superspace requires
that such an action could depend only upon $\bL_{SW}=\S dz$ and not
${\bar \bL}_{SW} = \bS d{\bar z}$. However to obtain a covariant action 
in six dimensions requires both $dz$ and $d{\bar z}$ to appear.


\appendix

In this appendix we will explicitly derive the values of the $J$ and $K$
integrals which appear above. If we define
$$
f = {R^2\Lambda^4 {\lambda}^2_{z}\partial_{\bz} \bs\over 
1 + R^2\Lambda^4|\partial_zs|^2}\ ,
\eqn\fdef
$$
then the integrals in question become (for simplicity we only consider the
lowest component terms)
$$
J = \int_{\Sigma} \partial_{\bz}{\bar f}d\bz \wedge {\lambda}\ ,\quad
K = \int_{\Sigma} \partial_{\bz}{f}d\bz \wedge {\lambda}\ .
\eqn\JKdef
$$
It is important to note that the function $f$ has singularities at the
roots $e_i$ of $Q$. Thus to evaluate $J$ and $K$ we must first cut holes
of radius $\epsilon$ around each of the $e_i$, $i=1,2,3,4$, forming a
new Riemann Surface $\Sigma_{\epsilon}$ with a boundary. The actual values 
for $J$ and $K$ are then found by taking the limit $\epsilon\rightarrow 0$.
Since ${\lambda}$ is a closed form the integrands in \JKdef\ can
be expressed as $d({\bar f}{\lambda})$ and 
$d({f}{\lambda})$ respectively. The surface integrals
can then be reduced to an integral around the boundary 
$\partial\Sigma_{\epsilon}$. Thus we find
$$
J = \sum_{i}\oint_{\gamma_i} d{z} {\bar f}{\lambda}_{z},\quad
K = \sum_{i}\oint_{\gamma_i}d{z} {f}{\lambda}_{z}\ ,
\eqn\JKeval
$$
where $\gamma_i$ is the $i$th component of $\partial\Sigma_{\epsilon}$.
If we substitute in the expression for $f$ and expand in powers of $\epsilon$
we find
$$
J = -{1\over 2}\sum_{i}\oint_{\gamma_i}d{z}\ 
{1\over \bz}{1\over Q}{1\over {\bar Q}} + {\cal O}(\epsilon),\quad
K = -{1\over 2}\sum_{i}\oint_{\gamma_i}d{z}\  
{1\over {z}}{1\over {Q}^2} + {\cal O}(\epsilon)\ .
\eqn\JKevaltwo
$$
From here it is easy to see that the integrand of $J$ has no simple
poles in ${z}$ and so $J=0$. However $K$ does have simple poles. 
After taking the $\epsilon \rightarrow 0$ limit we find
$$
K = -\pi i \sum_i {1\over {e}_i} 
\prod_{j\ne i}{1\over {e}_i-{e}_j}\ ,
\eqn\Keval
$$
in fact one can show that all the $\epsilon$ dependent terms vanish.
where we have used the  values  $e_i=\pm\sqrt{u\pm\Lambda^2}$ of the
four roots.
Note that despite its complicated  definition, $K$ 
depends holomorphically on $u$ and is independent
of the constant $R^2\Lambda^4$.
Since the four roots $e_i$ are given by $\pm\sqrt{u\pm\Lambda^2}$ we obtain
$$
K ={ \pi i\over {u}^2 -\Lambda^4}\ .
\eqn\Kevaltwo
$$
This is a remarkably simple expression and up to a constant it is the inverse
of the discriminant of $\Sigma$. Lastly we wish to show that $K$ is actually
equal to 
$$
E = -{d{\tau}\over d{u}}\left({d{a}\over d{u}}\right)^2\ .
\eqn\Edef
$$
To this end it can be shown that both $K$ and $E$  are modular functions
of weight zero. Thus $K^{-1}E$ is also a modular function of weight zero. 
Clearly  $K$  has simple poles at $u=\pm\Lambda^2$ and  furthermore, as 
is well known
[\SW], $\tau$ and $a$ also have singularities only at these points.
From the known solution [\SW] one
can explicitly check that $E$ has only simple poles at $u=\pm\Lambda^2$.
Hence $K^{-1}E$ is non-singular and thus must be 
constant. Finally, by examining the 
perturbative regime where ${u}\rightarrow \infty$, $a = -2\pi i \sqrt{u}$ 
and $\tau = \tau_0 + {i\over \pi}{\rm ln}(u/\Lambda^2)$, one can see that 
indeed $K=E$.

\refout

\end